\documentclass[%
 ,twocolumn%
 ,secnumarabic%
,amssymb, amsmath,nobibnotes, aps, prl]{revtex4}
\usepackage[dvipdfm]{graphicx}
\usepackage{latexsym,mathrsfs}
\usepackage{dcolumn}% Align table columns on decimal point
\usepackage{bm}% bold math

\def \ve#1{\mbox{\boldmath $#1$}}

\begin{document}

\title{Universal structure of two and three dimensional self-gravitating systems in the quasi-equilibrium state}

\author{Tohru Tashiro}
\affiliation{Department of Physics, Ochanomizu University, 2-1-1 Ohtuka, Bunkyo, Tokyo 112-8610, Japan}

\date{\today}

\begin{abstract}
We study a universal structure of two and three dimensional self-gravitating systems in the quasi-equilibrium state. It is shown numerically that the two dimensional self-gravitating system in the quasi-equilibrium state has the same kind of density profile as the three dimensional one.
We develop a phenomenological model to describe this universal structure by using a special Langevin equation with a distinctive random noise to self-gravitating systems.
We find that the density profile derived theoretically is consistent well with results of observations and simulations. 
\end{abstract}

\maketitle

\section{Introduction}

%Long range interacting systems%
Systems with long range forces exhibit various specific properties which systems with short range forces do not have. 
The prime example of them is the presence of another stable state differing from the thermal equilibrium state.
In this paper, we call it {\it quasi-equilibrium state} (QES).
It has been found numerically  that the distribution of QES depends strongly on initial conditions whereas the thermal equilibrium is independent on initial states.
QES of Hamiltonian mean field model which is a toy model of systems with long range interactions varies from a ferromagnetic state to a paramagnetic (homogeneous) state with increase of the initial total energy \cite{Levin14}.
We have obtained numerically the result that the number density $\mbox{$\mathcal{N}$}$ of three dimensional self-gravitating system (3DSGS) in the real space can be approximated by the following representation when the initial virial ratio $V(\equiv2K/\Omega)$ is 0 where $K$ and $\Omega$ are the total kinetic and gravitational energy respectively \cite{Tashiro10,Tashiro11}: 
\begin{equation}
\mbox{$\mathcal{N}$}(r) \simeq \frac{\mbox{$\mathcal{N}$}(0)}{(1+r^2/a^2)^\kappa}
\label{eq:fitf}
\end{equation}
with $\kappa \sim 3/2$, where $r$ means the distance from the center of system.
The number density of globular clusters consisting of about hundreds of thousands of stars which is the best example of 3DSGS also can be depicted by Eq.~(\ref{eq:fitf}) with $\kappa \sim 3/2$ \cite{Binney08,Peterson75,Trager95}: This density profile is universal for 3DSGS.

The recent observations have made it clear that the universality is not just limited to 3DSGS.
As will be discussed later, the cylindrical symmetric filamentary structure of molecular clouds can be treated as a two dimensional self-gravitating system (2DSGS). 
The {\em Herschel} Space Observatory revealed the 27 filamentary structures in IC~5146 which is a reflection nebula in the Cygnus and the number density of molecular clouds have a cylindrical symmetry around the axis of filament \cite{Arzoumanian11}.
All of the number densities of molecular clouds around the axis of filament can be fitted by Eq.~(\ref{eq:fitf}) with $\kappa$ from 0.75 to 1.25 \cite{Arzoumanian11}, where $r$ means the distance from the axis of filament.
 On the other hand, the equation (\ref{eq:fitf}) with $\kappa=1$ was utilized in order to describe the number density of a filament in the Taurus \cite{Stepnik03}. 
So, 2DSGS has a similar universality depicted by Eq.~(\ref{eq:fitf}) with $\kappa\sim1$.
%The interaction potential energies of 2D and 3DSGS have a difference in whether the potential is bounded or not.
The interaction potential energies of 2D and 3DSGS have a difference between being bounded and unbounded which will be explained later.
However, the universality is beyond the difference. 

Obviously, the universal number density cannot be derived by assuming the system is in the isothermal equilibrium: The equilibrium state for 2DSGS has an exact solution for number density represented by Eq.~(\ref{eq:fitf}) with $\kappa=2$ \cite{Ostriker64,Sire02}. On the other hand, the number density of 3DSGS in the thermal equilibrium decays with $r^{-2}$ at large radius $r$ \cite{Binney08,Chavanis02}.
Therefore, a new theory explaining the physical mechanism behind the universality is necessary.

In this paper, we shall derive the universal number density Eq.~(\ref{eq:fitf}) from one phenomenological model particular for gravity by utilizing a special Langevin equation for QES and corresponding Fokker-Planck equation in $\mu$ dimensional space where $\mu$ is 2 or 3. We shall treat them in the over-damped limit.
Indeed, the {\em normal} Langevin equation in the limit and the corresponding Fokker-Planck equation (i.e., Smoluchowski equation) is appropriate for describing 3DSGS enclosed in a box near the isothermal equilibrium \cite{Chavanis02}, since the Maxwell-Boltzmann distribution is stable for 3DSGS in the thermal equilibrium state with the total mass and energy fixed if the radius of system is less than a critical value \cite{Antonov62,Lynden-Bell68}.
However, it is well-known empirically that the SGSs without boundary are trapped by another stable state, that is QES, so that we must modify the Langevin and the Fokker-Planck equation in order to describe the state.
In particular, we shall introduce another noise whose foundation is so simple.

This paper is organized as follows. Firstly, we review the two dimensional gravity and show that the cylindrically symmetric 3DSGS is equivalent to 2DSGS mathematically . Moreover, we exhibit results of N-body simulations of it which are the same as observations of molecular clouds. Next, a Fokker-Planck equation to describe the universal number density is derived. In doing so, we start a Langevin equation specialized for gravity. Lastly, numerical solutions of the Fokker-Planck equation are shown and we reveal that the results fit well with observations and numerical simulations.  

\section{Two dimensional gravity and N-body simulations of two dimensional SGS}
\subsection{Two dimensional gravity}

The two dimensional gravitational potential per unit mass $\phi^2(\ve{r})$ generated by mass source $\rho^2(\ve{r})$ satisfies the following Poisson equation in the two dimensional space. 
\begin{equation}
\frac{1}{r}\frac{\partial}{\partial r}\left\{r\frac{\partial\phi^2(\ve{r})}{\partial r}\right\}
 + \frac{1}{r^2}\frac{\partial^2\phi^2(\ve{r})}{\partial \theta^2}=4\pi{G}'\rho^2(\ve{r})
\end{equation}
where the subscript 2 means the two dimensional space and we denote the two dimensional gravitational constant by $G'$ in order to distinguish it from the {\it ordinary} gravitational constant $G$.
If the distribution of mass is circular symmetric, the above equation becomes
\begin{equation}
\frac{1}{r}\frac{{\rm d}}{{\rm d}r}\left\{r\frac{{\rm d}\phi^2(r)}{{\rm d}r}\right\}
 =4\pi{G}'\rho^2({r}) \ .
\label{eq:2DPoisson}
\end{equation}

When the mass source is a mass point with a mass $m$ existing in the origin, $\rho^2({r})=m\delta(0)/\pi r$. Then, the Poisson equation is altered to
\begin{equation}
\frac{1}{r}\frac{{\rm d}}{{\rm d}r}\left\{r\frac{{\rm d}\phi^2(r)}{{\rm d}r}\right\}=\frac{4m{G}'\delta(0)}{r} \ .
\end{equation}
Therefore, the potential can be solved like
\begin{equation}
\phi^2(r) = 2{G}'m\ln r + \mbox{const.}
\label{eq:inter}
\end{equation}
Then, the interaction potential of 2DSGS is bounded whereas one of 3DSGS is proportional to $-1/r$.  

Finally, we shall provide a brief explanation for the reason that cylindrically symmetric filaments of molecular clouds can be regarded as 2DSGS:
By using the mass density of molecular clouds $\rho_{\rm mc}(r)$ and the gravitational potential per unit mass $\phi_{\rm mc}(r)$ where $r$ means a distance  from the axis, the Poisson equation becomes 
\begin{equation}
\frac{1}{r}\frac{{\rm d}}{{\rm d}r}\left\{r\frac{{\rm d}\phi_{\rm mc}(r)}{{\rm d}r}\right\}=4\pi G\rho_{\rm mc}(r) \ ,
\end{equation}
which is mathematically equivalent to the Poisson equation for 2DSGS Eq.~(\ref{eq:2DPoisson}).

\subsection{N-body simulations}

Because the equivalence is merely formal, we followed the time evolution of $10^4$-body system in the two dimensional space interacting by Eq.~(\ref{eq:inter}) numerically in order to investigate QES of 2DSGS.

A polytrope solution with a polytrope index $n$ was adopted as the initial condition. 
The solution with $n=\infty$ corresponds to the thermal equilibrium state, 
so  the finite $n$ represents the deviation from equilibrium. 
Generally, the solution in the three dimensional space is well-known \cite{Binney08}. Here, we have extended it to the two dimensional space: The distribution function $f(r,v)$ in phase space can be shown as
\begin{equation}
f(r,v) \propto \left\{
\begin{array}{cc}
\mbox{$\mathcal{E}$}^{{n}-1} & (\mbox{$\mathcal{E}$} < 0) \\
0                                          &  (\mbox{$\mathcal{E}$} \ge 0)
\end{array}
\right. 
\end{equation}
where $\mbox{$\mathcal{E}$} \equiv \phi^2(R) - \left\{\frac{1}{2}v^2+\phi^2(r)\right\}$ and $R$ means a radius of the system.
Then, we run the N-body simulation by varying an initial virial ratio for 2DSGS $V'$ which is different from a virial ratio for 3DSGS (See Refs.~\cite{Chavanis06} and \cite{Teles10}). Note that the system is not enclosed into a circle.

\begin{figure}[h]
  \begin{center}
%    \begin{minipage}{14pc}
%      \hspace{-3pc}
      \includegraphics[scale=0.65]{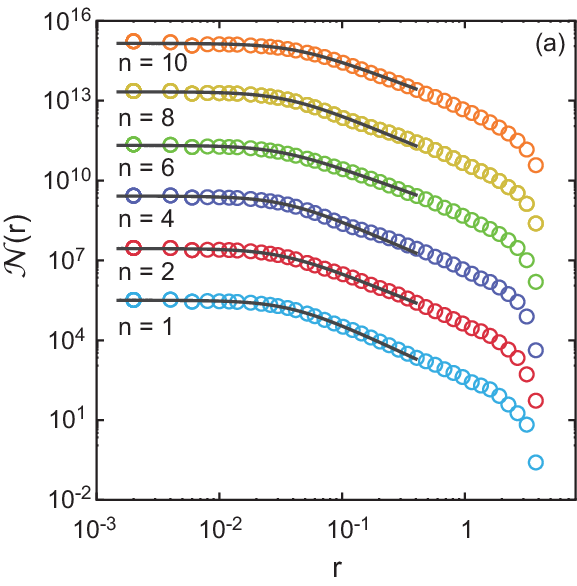}
 %   \end{minipage}
 %   \hspace{5pc}%
 %   \begin{minipage}{14pc}
 %     \hspace{-2pc}
      \includegraphics[scale=0.65]{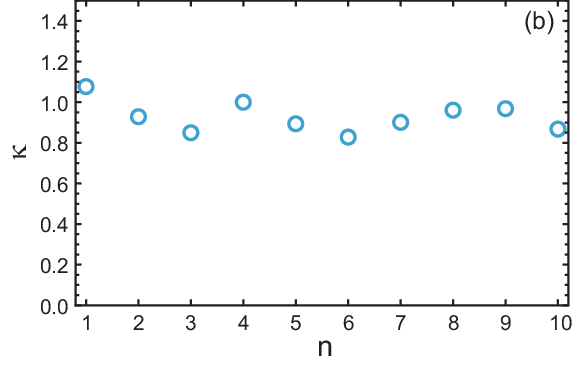}
 %   \end{minipage} 
    \caption{\label{fig:numberandkappa}(color online) (a) Number densities at QES for several initial polytrope indexes are plotted by open-circles. The curve passing the circles is a fitting function, Eq.~(\ref{eq:fitf}). 
So as to be seen easily, each density is shifted by two digits to the below. (b) The optimum values of $\kappa$ in Eq.~(\ref{eq:fitf}) for fitting  the number densities vs the initial polytope index $n$.
}
  \end{center}
\end{figure}

Specifically when $V'=0$, it is found that the number density in QES has a universality and the density around the center of the system can be fitted well by Eq.~(\ref{eq:fitf}).
The results are shown in Fig.~\ref{fig:numberandkappa}. It can be seen in Fig.~\ref{fig:numberandkappa}(a) that the number density from $n=1$ to 10 has the same profile.
In addition, figure~\ref{fig:numberandkappa}(b) denotes that $\kappa$ which is the index in Eq.~(\ref{eq:fitf}) ranges from 0.8 to 1.1. These results have a good consistency with the observations of molecular clouds \cite{Arzoumanian11,Stepnik03}.

\section{Fokker-Planck model}
\label{sec:anoFPM}

We have made sure of a fact that the number density of 2DSGS without boundary in QES can be represented well by Eq.~(\ref{eq:fitf}) with $\kappa\sim 1$.
On the other hand, the number density of most globular clusters is well-known to be fitted by Eq.~(\ref{eq:fitf}) with $\kappa\sim3/2$ \cite{Binney08,Peterson75,Trager95}.
Furthermore, we have reported that the same density profile is obtained through N-body simulations of 3DSGS \cite{Tashiro10,Tashiro11}.
Therefore, we can conclude that the SGSs without boundary have the universal density profile depicted by Eq.~(\ref{eq:fitf}) in QES.

Here, let us derive the density profile uniformly by a special Fokker-Planck equation in the $\mu$ dimensional space where $\mu$ is 2 or 3 \footnote{
A reason why we adopt the Fokker-Planck equation approach rather than the Boltzmann equation is as follows:
It is founded numerically that 2D and 3DSGS reach the universal quasi-equilibrium state specifically when initial virial ratios are 0. The initial conditions cause the {\em cold collapse} resulting in the high density around the center of the system. Therefore, each collision there cannot be distinguished and the Boltzmann equation premising that the collision is distinguishable  is not appropriate. 
}. Before constructing the Fokker-Planck equation, we shall model forces influencing an element of the system. That is, we shall begin by constructing a Langevin equation.

We assume that the frictional force $-m\gamma\dot{\ve{r}}^\mu(t)$ and the random noise with constant intensity $\sqrt{2D}\ve{\xi}^\mu(t)$ which are essential for a many-body system to reach the thermal equilibrium state act on the element \cite{Chavanis02}, where $m$ is a mass of the element, $\ve{r}^\mu$ is a position vector, and $\ve{\xi}^\mu(t)$ means Gaussian-white noises \footnote{With regard to $\mu=3$, the friction from SGS and the time at which the orbit of the element becomes stochastic are known as the {\em dynamical friction} \cite{Chandra43} and the {\em two body relaxation time} \cite{Spitzer87}, respectively. However, no theory for the two dimensional space has been done.}. 
The index represents that the vector is in the $\mu$ dimensional space.
Postulating the system to be circular ($\mu=2$) or spherical symmetric ($\mu=3$), 
one can see that the element is also influenced by a mean gravitational force $-F^\mu(r)$ along the radial direction of the system which is derived by differentiating $m\phi^\mu(r)$: $-F^\mu(r) = -m\partial_r\phi^\mu(r)$ where $\phi^\mu(r)$ is the mean gravitational potential in $\mu$ dimensional space and $r\equiv|\ve{r}^\mu|$ in which the index $\mu$ are abbreviated for simplify.
However, this is just a mean gravity. It is natural to consider that the element actually is influenced by a fluctuating gravity around the mean value: The number density producing $\phi^\mu(r)$ through the Poisson equation is the mean value, and the actual distribution of elements must fluctuate around the value.  
This means that another noise which prevents the system from reaching the thermal equilibrium state is added to the normal Langevin equation, and so this system goes to another stable state, that is QES.
Therefore, we can consider the noise distinctive to SGS.

If assuming the intensity of the noise to be constant, we can obtain the following Langevin equation:
\begin{align}
\lefteqn{m\ddot{\ve{r}}^\mu(t) + m\gamma\dot{\ve{r}}^\mu(t)} \nonumber \\
&= -F^\mu(r)\left\{1+\sqrt{2\epsilon}{\eta}(t)\right\}\ve{e}^\mu_r + 
 \sqrt{2D}\ve{\xi}^\mu(t)
\label{eq:lan1}
\end{align}
where $\ve{e}^\mu_r$ is a unit vector along the radial direction in $\mu$ dimensional space.
In the over-damped limit, equation~(\ref{eq:lan1}) becomes 
\begin{align}
m\gamma\dot{\ve{r}}&^\mu(t)
=-F^\mu(r)\left\{1+\sqrt{2\epsilon}{\eta}(t)\right\}\ve{e}^\mu_r \nonumber \\
& -\frac{\partial}{\partial r}\frac{\epsilon}{2m\gamma}F^\mu(r)^2\ve{e}^\mu_r
+ \sqrt{2D}\ve{\xi}^\mu(t) \ .
\end{align}
The second term on the right hand side of the above equation is a correction term in order to regard products as the Storatonovich product \cite{Sekimoto99}D

The corresponding Fokker-Planck equation is given by
\begin{align}
\frac{\partial}{\partial t}&P^\mu(r,t) = \frac{D}{(m\gamma)^2}
\left\{\frac{\partial^2}{\partial r^2}
+\frac{\mu-1}{r}\frac{\partial}{\partial r}\right\}P^\mu(r,t) \nonumber \\
& \mbox{} + \frac{1}{m\gamma}\frac{1}{r}
\frac{\partial}{\partial r}rF^\mu(r)P^\mu(r,t) \nonumber \\
&+\frac{\epsilon}{(m\gamma)^2}\left\{
\frac{\partial^2}{\partial r^2}+\frac{\mu-1}{r}\frac{\partial}{\partial r}
+\frac{\mu-1}{r^{\mu-1}}\frac{\partial}{\partial r}r^{\mu-2} 
\right\} \nonumber \\
&\times F^\mu(r)^2P^\mu(r,t) \ .
\end{align}
We are treating the system as a circular or a spherical symmetric one including $N$ elements. 
So, PDF $P^\mu$ is a function of the distance from the origin $r$.
Note that a relation among PDF $P^\mu$, the number density $\mbox{$\mathcal{N}$}^\mu$ and the mass density $\rho^\mu$ is as follows: $P^\mu = \mbox{$\mathcal{N}$}^\mu/N = \rho^\mu/(mN)$.

Let us use $\mbox{$\mathcal{P}$}^\mu(r,t) = J^\mu(r)P^\mu(r,t)$ instead of $P^\mu$, where 
$J^\mu$ means Jacobian determinant, $J^2 = 2\pi r$ and $J^3 = 4\pi r^2$: It can be depicting by $J^\mu=2^{\mu-1}\pi r^{\mu-1}$. In doing so, we can obtain
\begin{align}
&\frac{\partial}{\partial t}\mbox{$\mathcal{P}$}^\mu(r,t) = \frac{D}{(m\gamma)^2}
\left\{\frac{\partial^2}{\partial r^2}-\frac{\partial}{\partial r}\frac{\mu-1}{r}\right\}\mbox{$\mathcal{P}$}^\mu(r,t) \nonumber \\
&\hspace{-0ex} + \frac{1}{m\gamma}\frac{\partial}{\partial r}F^\mu(r)\mbox{$\mathcal{P}$}^\mu(r,t)
+\frac{\epsilon}{(m\gamma)^2}\frac{\partial^2}{\partial r^2}F^\mu(r)^2\mbox{$\mathcal{P}$}^\mu(r,t) \ .
\end{align}

When the system reaches QES, $\partial_t\mbox{$\mathcal{P}$}_{\rm qe}^\mu(r)=0$. Here, we integrate the Fokker-Planck equation by $r$. Owing to use $\mbox{$\mathcal{P}$}_{\rm qe}^\mu$, the integration becomes easier:
\begin{align}
&\left\{\frac{D}{(m\gamma)^2}+\frac{\epsilon F^\mu(r)^2}{(m\gamma)^2}\right\}
{\mbox{$\mathcal{P}$}^\mu_{\rm qe}}'(r) \nonumber \\
&-\left[\frac{D}{(m\gamma)^2}\frac{\mu-1}{r} 
- \frac{2\epsilon F^\mu(r){F^\mu}'(r)}{(m\gamma)^2} - \frac{F^\mu(r)}{m\gamma}\right]\mbox{$\mathcal{P}$}^\mu_{\rm qe}(r)  \nonumber \\
&= \mbox{const.}
\label{eq:st1}
\end{align}
Since $P_{\rm qe}(0)$ and $P_{\rm qe}'(0)$ are bounded, $\mbox{$\mathcal{P}$}^\mu_{\rm qe}(0)={\mbox{$\mathcal{P}$}^\mu_{\rm qe}}'(0)=0$.  Therefore, the constant of the right hand side of Eq.~(\ref{eq:st1}) becomes 0:
\begin{equation}
{\mbox{$\mathcal{P}$}^\mu_{\rm qe}}'(r) = -\frac
{rF^\mu(r)\left\{m\gamma+2\epsilon {F^\mu}'(r)\right\}-(\mu-1)D}
{r\left\{{D}+\epsilon F^\mu(r)^2\right\}}\mbox{$\mathcal{P}$}^\mu_{\rm qe}(r) \ .
\end{equation}
If using the number density in QES $\mbox{$\mathcal{N}$}^\mu_{\rm qe} (= N\mbox{$\mathcal{P}$}^\mu_{\rm qe}/J^\mu)$, we can obtain
\begin{align}
\lefteqn{{\mbox{$\mathcal{N}$}^\mu_{\rm qe}}'(r) = } \nonumber \\
&\hspace{-0ex}-\frac
{rF^\mu(r)\left\{m\gamma+2\epsilon {F^\mu}'(r)\right\}+(\mu-1)\epsilon F^\mu(r)^2}
{r\left\{{D}+\epsilon F^\mu(r)^2\right\}}\mbox{$\mathcal{N}$}^\mu_{\rm qe}(r) \ .
\label{eq:unit1}
\end{align}

By substituting the gravitational potential per unit mass $\phi^\mu(r)(= \frac{1}{m}\int{\rm d}rF^\mu(r))$ into the Poisson equation $\mbox{$\bigtriangleup$}\phi^\mu=4\pi \mbox{$\mathcal{G}$}^\mu\rho^\mu_{\rm qe}={4\pi \mbox{$\mathcal{G}$}^\mu m\mbox{$\mathcal{N}$}^\mu_{\rm qe}(r)}$, an equation governing $F^\mu$ can be obtained as follows:
\begin{equation}
{F^\mu}'(r)+\frac{\mu-1}{r}F^\mu(r) ={4\pi \mbox{$\mathcal{G}$}^\mu m^2\mbox{$\mathcal{N}$}^\mu_{\rm qe}(r)} \ ,
\label{eq:unit2}
\end{equation}
where $\rho^\mu_{\rm qe} = m\mbox{$\mathcal{N}$}^\mu_{\rm qe} = mNP^\mu_{\rm qe} = mN\mbox{$\mathcal{P}$}^\mu_{\rm qe}/J^\mu$ and $\mbox{$\mathcal{G}$}^\mu$ relates quantities appeared previously as $\mbox{$\mathcal{G}$}^2={G}'$ and $\mbox{$\mathcal{G}$}^3={G}$. 

Here, we nondimensionalize these equations by using the following units of length and force:
\begin{equation}
\textstyle{\rm [length]} = \sqrt{\frac{\mu^2(\mu+2)D}{2\pi{\cal{G}}^\mu m^2{\cal{N}}^\mu_{\rm qe}(0)\left\{\mu^2m\gamma+8\pi(\mu^2+4\mu+2)\epsilon{\cal{G}}^\mu m^2{\cal{N}}^\mu_{\rm qe}(0)\right\}}}
\end{equation}
\begin{equation}
\textstyle
{\rm [force]} = \sqrt{\frac{8\pi\mu^2(\mu+2)D{\cal{G}}^\mu m^2{\cal{N}}^\mu_{\rm qe}(0)}{\mu^2m\gamma+8\pi(\mu^2+4\mu+2)\epsilon{\cal{G}}^\mu m^2{\cal{N}}^\mu_{\rm qe}(0)}}
\end{equation}
Then, equations~(\ref{eq:unit1}) and (\ref{eq:unit2}) are altered to
\begin{align}
\textstyle
\lefteqn{{{{\bar{\mbox{$\mathcal{N}$}}}^\mu}_{\rm qe}}{}'(\bar{r})=} \nonumber \\
&\hspace{-3ex}-2\mu(\mu+2)\frac{\bar{r}\bar{F}^\mu(\bar{r})\left\{1+2\mu q{\bar{F}^\mu}{}'(\bar{r})\right\}+\mu(\mu-1)q\bar{F}^\mu(\bar{r})^2}{\bar{r}\left\{\mu+2(\mu^2+4\mu+2)q+2\mu^2(\mu+2)q\bar{F}^\mu(\bar{r})^2\right\}} \nonumber \\
&\times\bar{\mbox{$\mathcal{N}$}}^\mu_{\rm qe}(\bar{r}) \ ,
\label{eq:sys1}
\end{align}
and
\begin{equation}
\bar{F}^\mu{}'(\bar{r})+\frac{\mu-1}{\bar{r}}\bar{F}^\mu(\bar{r}) =\bar{\mbox{$\mathcal{N}$}}^\mu_{\rm qe}(\bar{r}) \ ,
\label{eq:sys2}
\end{equation}
where $q\equiv 4\pi\epsilon \mbox{$\mathcal{G}$}^\mu m^2\mbox{$\mathcal{N}$}^\mu_{\rm qe}(0)/(\mu m\gamma)$ and variables with overbars denote dimensionless. We should solve these equations with boundary condition $\bar{\mbox{$\mathcal{N}$}}^\mu_{\rm qe}(0)=1$.

\section{Results}
\label{results}

\begin{figure}[h]
  \begin{center}
 %   \begin{minipage}{14pc}
 %     \hspace{-3pc}
      \includegraphics[scale=0.75]{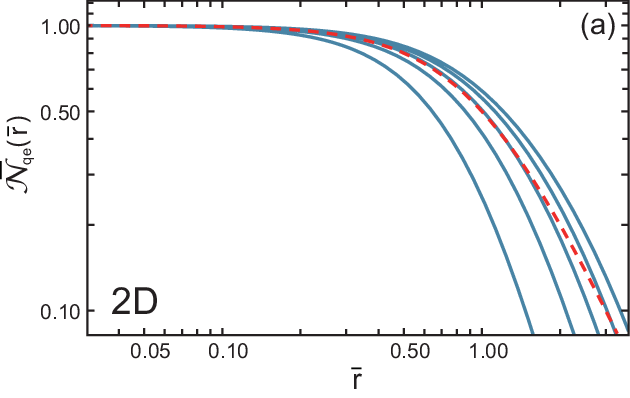}
 %   \end{minipage}
  %\vspace{10pc}%
  %  \begin{minipage}{14pc}
   %   \hspace{-2pc}
      \includegraphics[scale=0.75]{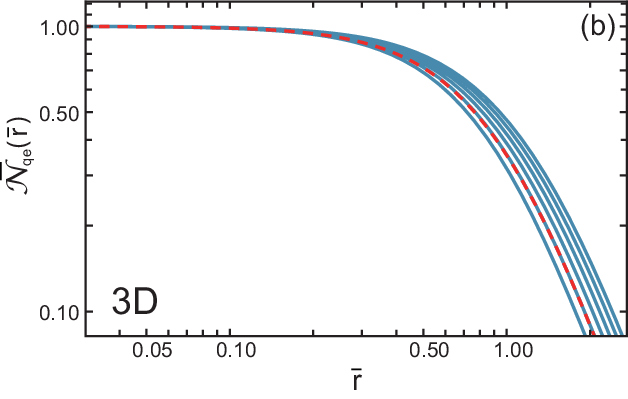}
   % \end{minipage} 
    \caption{\label{fig:result_mymodel}(color online) Numerical solutions of Eqs.~(\ref{eq:sys1}) and (\ref{eq:sys2}) for $\mu=2$ (a) and $\mu=3$ (b). As the curve changes from the left to the right in (a), $q$ gets larger from 0 to 0.28 in steps of 0.07. On the other hand, $q$ gets larger from 0 to 0.05 in steps of 0.01 in (b). The (red) dashed curve in (a) and (b) means $1/(1+\bar{r}^2)$ and $1/(1+\bar{r}^2)^{3/2}$, respectively. }
  \end{center}
\end{figure}

The numerical solutions for $\mu=2$ and $3$ are shown in Fig.~\ref{fig:result_mymodel} by changing $q$.
The curves with $q=0$ on both figures correspond to the thermal equilibrium state.
For comparison with observations, $1/(1+\bar{r}^2)$ and $1/(1+\bar{r}^2)^{3/2}$ are also plotted by a dashed curve in Fig.~\ref{fig:result_mymodel} (a) and (b), respectively. 
The dashed curves are typical best-fit ones for densities of molecular clouds  or globular clusters. 
From Fig.~\ref{fig:result_mymodel} (b), one can see that the numerical result of our model with $q=0.01$ completely coincide with the typical number density.
Fig.~\ref{fig:result_mymodel} (a) also shows the good agreement of our model with observations and numerical simulations for small radius by setting $q=0.56$, although the deviation between two curves increases as $r$ gets larger.
Therefore, we can understand that the best-fit curves are derived from our model by varying $q$ appropriately. 

Finally, we examine the range of the index $\kappa$.
From Eqs.~(\ref{eq:sys1}) and (\ref{eq:sys2}),
the number density around the center of the system in QES can be described by
\begin{equation}
\bar{\mbox{$\mathcal{N}$}}^\mu_{\rm qe}(\bar{r}) \simeq \frac{1}{(1+\bar{r}^2)^{\kappa(q)}}
\label{eq:fitf2}
\end{equation}
where $\kappa$ is a function of $q$:
\begin{equation}
\kappa(q) = \frac{(\mu+2)\left\{1+(\mu+1)q\right\}}{\mu+2(\mu^2+4\mu+2)q} \ .
\end{equation}
From this equation, because of $q\ge0$, one can see the range of $\kappa$  as follows:
\begin{equation}
\frac{(\mu+1)(\mu+2)}{2(\mu^2+4\mu+2)}<\kappa
\le\frac{\mu+2}{\mu} \ .
\end{equation}
Therefore, with regard to $\mu=2$ and $\mu=3$, $\frac{3}{7}<\kappa \le 2$ and $\frac{10}{23}<\kappa \le \frac{5}{3}$, respectively.  Both ranges include the observed indexes.
However, the observed ranges of the index are so narrower, which means that the value of $q$ is limited. 
The limited $q$ can be regarded as a new kind of {fluctuation-dissipation relation} \cite{Kubo91} which is particular for SGS because $q$ includes a ratio of the intensity of {\em mean gravity fluctuation} $\epsilon$ to the friction coefficient $\gamma$.
Indeed, this phenomenological approach cannot provide an answer to a question why the specific value of $q$ is selected.

\section{Concluding remarks}
 
In this paper, we made it clear numerically that 2DSGS without boundary goes to QES in which the density profile is depicted by Eq.~(\ref{eq:fitf}) with $\kappa\sim1$. It is well-known that QES of 3DSGS without boundary can be described by the same one through the observations of globular clusters and N-body simulations. That is, it was shown that there is the universal QES for the SGSs:
As mentioned before, the interaction potential at 2DSGS are qualitatively different from one at 3DSGS.
Moreover, the isothermal equilibrium of SGS with boundary differs between 2D and 3D \cite{Chavanis02,Sire02,Kiessling89,Aly94}.
On the contrary, the paper unveiled that 2D and 3DSGS without boundary have the same universality in QES.
Furthermore, we developed the model to derive the universal density profile from the special Langevin equation including the distinctive noise of SGS.
Indeed, the solution of the corresponding Fokker-Planck equation in QES was depicted by Eq.~(\ref{eq:fitf}).
In addition, it was found that the index $\kappa$ in Eq.~(\ref{eq:fitf}) can be represented as a function of the intensity of the particular noise for SGS, the friction coefficient, and others. So,  we showed the range of $\kappa$ for each dimension analytically which has a good consistency with observations and simulations. 

The theory justifying the frictional force and the random noise is indispensable for constructing the Langevin equation. As mentioned before, there are theories for 3DSGS \cite{Chandra43,Spitzer87}, but not for 2DSGS.
Therefore, such a theory must be developed.

\section{Acknowledgments}
We would like to thank Prof. Takayuki Tatekawa for N-body simulations and members of astrophysics laboratory at Ochanomizu
University for extensive discussions.
This work was supported by JSPS Grant-in-Aid for Challenging Exploratory Research 24654120.

%%%%%%%%%%%%%%%%% BIBLIOGRAPHY IN THE LaTeX file !!!!! %%%%%%%%%%%%%%%%%%%%%%

\end{document}